\documentclass[aps,reprint,superscriptaddress,longbibliography,floatfix]{revtex4-1}
\usepackage{graphicx}
\usepackage{epstopdf}
\usepackage{bm}
\usepackage{subfigure}
\usepackage{sidecap}

\usepackage{graphicx}
\usepackage{color}
\usepackage{epstopdf}
\usepackage{amssymb}
\usepackage{amsmath}

\usepackage{bm}
\graphicspath{{Figures/}}
\usepackage{subfigure}
\usepackage{sidecap}
\usepackage {times}

\newcommand{\si}{\sigma}
\newcommand{\al}{\alpha}

\newcommand{\eps}{\epsilon}

\newcommand{\lam}{\lambda}

\newcommand{\De}{\Delta}

\newcommand{\bsig}{{\bm\sigma}}
\newcommand{\bdel}{{\bm\delta}}

\newcommand{\be}{\begin{equation}}
\newcommand{\ee}{\end{equation}}

\newcommand{\bea}{\begin{eqnarray}}
\newcommand{\eea}{\end{eqnarray}}
\newcommand{\bd}{\begin{displaymath}}
\newcommand{\ed}{\end{displaymath}}
\newcommand{\ba}{\begin{array}}
\newcommand{\ea}{\end{array}}
\newcommand{\bi}{\begin{itemize}}
\newcommand{\ei}{\end{itemize}}
\newcommand{\bc}{\begin{center}}
\newcommand{\ec}{\end{center}}
\newcommand{\bfl}{\begin{flushleft}}
\newcommand{\efl}{\end{flushleft}}
\newcommand{\bfr}{\begin{flushright}}
\newcommand{\efr}{\end{flushright}}

\newcommand{\bl}{\begin{aligned}}
\newcommand{\el}{\end{aligned}}

\newcommand{\hh}{\hat{h}}
\newcommand{\hG}{\hat{G}}

\newcommand{\hep}{\hat{\epsilon}}

\newcommand{\tE}{\tilde{E}}
\newcommand{\tep}{\tilde{\epsilon}}

\newcommand{\tga}{\tilde{\gamma}}

\newcommand{\fs}{\frac{1}{2}}

\newcommand{\om}{i\omega_n}

\newcommand{\ra}{\rangle}
\newcommand{\la}{\langle}

\newcommand{\CR}{CeRh$_2$As$_2$}
\newcommand{\GR}{GdRu$_2$Si$_2$}

\def\br{{\bf r}}\def\bR{{\bf R}} 
\def\bk{{\bf k}} \def\bK{{\bf K}} \def\bq{{\bf q}}

\def\bQ{{\bf Q}}   
   
 \def\bd{{\bf d}} \def\bS{{\bf S}} 
 \def\bS{{\bf S}}

\def\bs{{\bf s}}

\def\da{\downarrow} \def\ua{\uparrow}

\def\={\!\!\!&=&\!\!\!}
\def\+{\!\!\!&&\!\!\!+~}
\def\-{\!\!\!&&\!\!\!-~}

\newcommand{\GRS}{GdRu$_2$Si$_2$}

\usepackage{color}
\usepackage[dvipsnames]{xcolor}
\usepackage{xcolor}

\usepackage[colorlinks=true,citecolor=blue]{hyperref}

\usepackage{hyperref,cleveref}

\begin{document}

\title{Local moment magnon spectrum in conduction electron tunnelling }

\author{Peter Thalmeier}  
\affiliation{Max Planck Institute for Chemical Physics of Solids, D-01187 Dresden, Germany}

\author{Alireza Akbari}
\affiliation{Beijing Institute of Mathematical Sciences and Applications (BIMSA), Huairou District, Beijing, 101408, China}

\date{\today}

\begin{abstract}
The surface tunnelling spectrum of a dual system consisting of localised moments with antiferromagnetic order 
coupled to conduction electrons by on-site exchange interaction is investigated. In the static approximation of
the local moment order it is known that magnetic band reconstruction leads to an anomalous tunnelling spectrum at the
magnetic ordering vector of local moments although the latter cannot contribute directly. In this work we 
consider dynamic effects by including the scattering of conduction electrons from the local moment magnon
excitations. They lead to self energies and renormalisation of conduction states which in turn appreciably
modify the tunnelling spectrum beyond the influence of static order, interpreted as the appearance of magnon sidebands.
\end{abstract}
\maketitle

\section{Introduction}
\label{sec:intro}

The tunnelling of conduction electrons in STM experiments provides various informations on the underlying electronic system: It gives an image of the topographic spectral electronic density at the surface by scanning the surface at fixed bias voltage. Complementary at a fixed position the change of bias voltage reveals the DOS function of conduction electrons. And  the combined surface scanning and voltage  change in quasiparticle interference (QPI) experiments can map the Fermi surface and dispersive features of the conduction bands.

When only the total charge current is measured and then one might expect
that magnetic signatures are not resolved. And yet this possibility has been demonstrated in tunnelling experiments  on the compound \GRS~\cite{spethmann:24,yasui:20} which has attracted some attention due to possibility of hosting a skyrmion phase \cite{wood:23}. It has dual electronic character with i) tightly bound 4f electrons in stable Gd$^{3+}$  $4f^7$  shells with large spin $S=7/2$ and orbital $L=0$ state. The resulting large moments $(g_J=2)$ order magnetically due to an RKKY type interaction mediated by ii) conduction electrons with Fermi surface nesting properties \cite{bouaziz:22,eremeev:23,matsuyama:23} that lead to an incommensurate helical order \cite{nomoto:20} of the localised Gd moments with wave vector $\bQ\simeq 0.4(\pi/a)$ oriented along a tetragonal in-plane axis of the \GRS~structure with $D_{4h}$ symmetry.

The f-electrons in the stable Gd$^{3+}$ shell will not contribute directly to the tunnelling structure. Surprisingly their magnetic order has nevertheless been identified in the STM images. In addition to the topographic peaks at tetragonal reciprocal lattice vectors \bK~the magnetic  order appears as  weaker satellites at $\bK\pm\bQ$ in the Fourier transformed scanning images.
It was argued in the theoretical investigation \cite{akbari:25} that this is due the reconstruction of conduction band states caused by the on-site exchange coupling to the molecular field of the Gd local moment structure. This produces an anomalous spectral density for the conduction electron charge that is seen in the tunnelling experiment, although the total  integrated charge density is unchanged. This anomalous density appears indeed at the $\pm\bQ$ satellite positions relative to the background electron density peaks at \bK.
This is an important observation since it provides the potential to investigate the static structure properties of (surface) magnetism structure from charge tunnelling experiments. It was also shown that the quasiparticle interference
due to surface impurities \cite{akbari:25} can give information of the reconstruction of the conduction electron bands e.g. by appearance of magnetic gaps. \\

These are all static pieces of information on the effect of local moment order to be obtained from STM. Now the interesting question arises to which extent the {\it dynamics} of local moment order, i.e. the magnon excitations might also leave a discernible fingerprint in the tunnelling {\it charge} current. In fact first experimental results  in this direction have been reported \cite{ozdemir:26}. This issue of possible dynamic magnon spectrum visibility in the charge STM tunnelling spectrum will be investigated in this work. It is a principal theoretical study and we do not intend to refer directly to \GRS~since the latter has an incommensurate helical Gd local moment structure where the calculation of the magnon spectrum itself would already be quite involved. 

We rather concentrate on a simplified model system with commensurate easy-axis (perpendicular to the surface) antiferromagnetic (AF) local moment order that couples to a single  2D tight binding conduction band corresponding to a near nesting condition for the   planar RKKY exchange.  The latter results from an onsite contact exchange interaction originating from the Schrieffer-Wolff mechanism.
 Aside from the static reconstruction of conduction bands by the AF order the excitation of virtual magnons leads to a conduction electron self energy that contains the information on the (surface) magnon spectrum. This self energy eventually appears as also in the tunnelling current spectrum and hence may be identified from its signatures. This might possibly provide a way to extract some features of the AF surface magnon spectrum from surface charge tunnelling experiments.\\

The work is organised as follows. In Sec.~\ref{sec:model} we define the demonstration model mentioned above in more detail, deriving briefly the magnon spectrum and their interaction with conduction electrons. Sec.~\ref{sec:spectrum} shows how the electron-magnon self energy appears in the tunnelling spectrum and Sec.~\ref{sec:selfenergy} provides the derivation of the self energy expressions and renormalized quasiparticle dispersion. The numerical results and their discussion are presented in Sec.~\ref{sec:numerical} and Sec.~\ref{sec:conclusion} gives the conclusions.
%
\begin{figure}
\includegraphics[width=0.99\columnwidth]{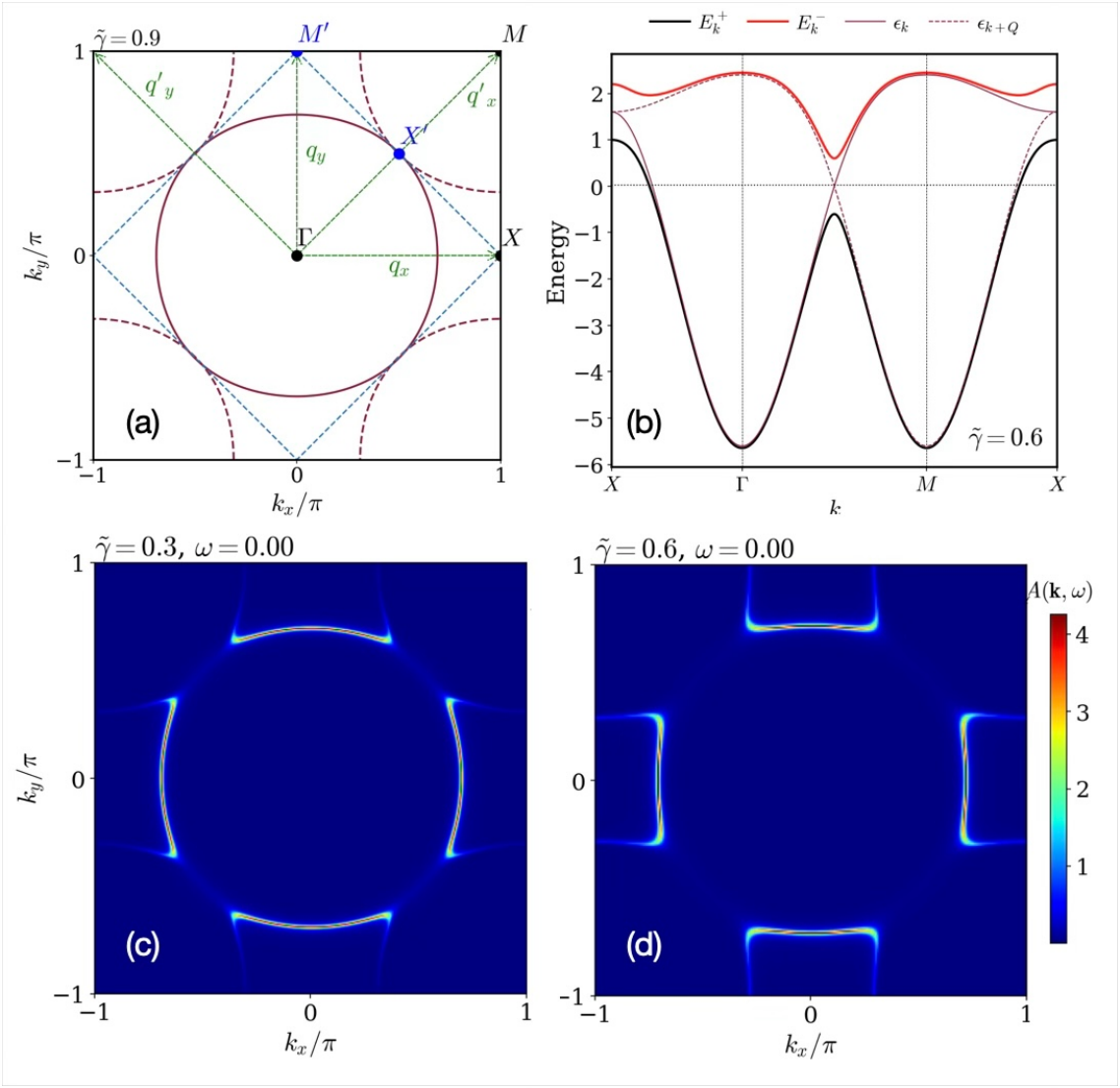}
\caption{(a) Fermi surface of bare $(\tga=0)$ conduction band $\epsilon_\bk$ ($\Gamma$-centered) and shifted
bands $\epsilon_{\bk+\bQ}$ (M-centered) which correspond to bare dispersion in (b). For finite $\tga$ an AF
gap opens at the crossing point X' at the Fermi energy $\epsilon_F$ leading to reconstructed bands and their spectral functions (c,d)
(for $\omega=\epsilon_F$). The original BZ  (momenta in units of $\pi/a$ with lattice constant $a\equiv 1$) 
and antiferromagnetic BZ (AFBZ) correspond  to full black and dashed blue lines in (a), with momentum coordinates ($q_x,q_y$) and ($q'_x,q'_y$), respectively, having
symmetry points $\Gamma$ (0,0), $X(\pi,0)$, $M(\pi,\pi)$ and $X'(\frac{\pi}{2},\frac{\pi}{2})$, $M'(0,\pi)$ as indicated. 
 Here $t=1,t'=0.4,\epsilon_F=0$ are used.}
\label{fig:FS-spec}
\end{figure}
%

\section{Two component model for localised moments and  conduction bands}
\label{sec:model}

\begin{figure}
\includegraphics[width=\columnwidth]{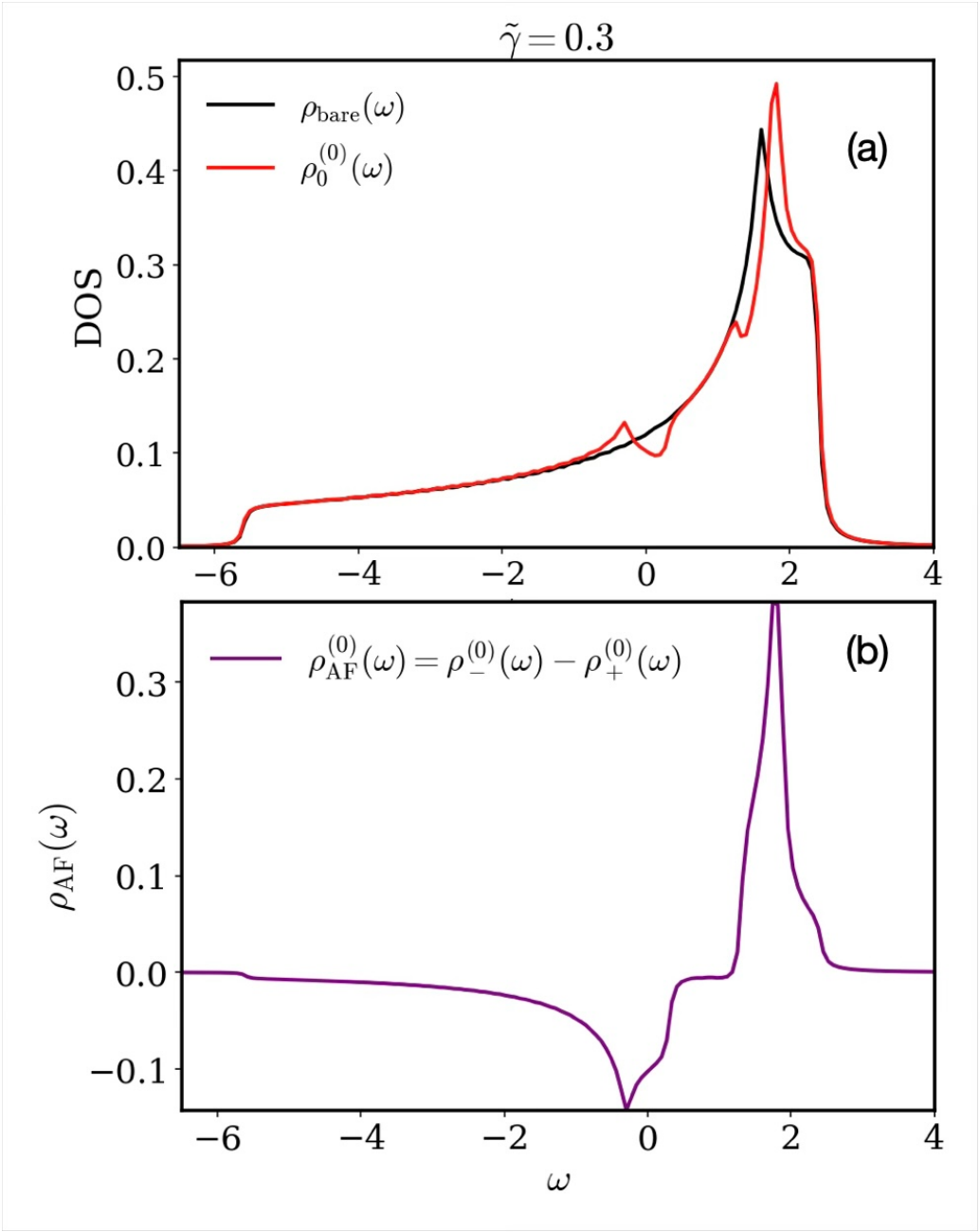}
\caption{(a) Normal \bk- integrated spectral function (DOS) in Eq.~(\ref{eq:DOSstatic})  for moderate $(\tga=0.3)$ exchange coupling including {\it only} static band reconstruction setting $\Sigma=0$.  For comparison the black curve corresponds to  bare  band spectrum $(\tga=0)$ equal to the conduction electron DOS with $N_0=\rho_{bare}(0)$ denoting its value at the Fermi level}. b) Anomalous spectral contribution in Eqs.~(\ref{eq:totDOS1},\ref{eq:DOSstatic}) induced at $\pm\bQ$ wave vectors by AF order parameter. For larger $\tga$ see Fig.~\ref{fig:fullspec}.  Energies and frequencies $\omega$ $(\hbar\equiv 1)$ are given in units of $t\equiv 1$ and DOS and spectral function in units of $1/t$. This applies also to all following figures.
\label{fig:disp-DOS}
\end{figure}

The model we consider consists of subsystems with localised  moment (pseudo-) spins $\bS_i$ originating e.g. from Kramers doublets of crystalline electric field (CEF)- split 4f states  and conduction electrons with band dispersion $\epsilon_\bk$ and spin density $\bs(\br)$ which are coupled by an on-site contact exchange interaction $I_{ex}(\br-\bR_i)=I_{ex}\delta(\br-\bR_i)$  that is generated from c-f hybridisation  and 4f-intrashell Coulomb interaction via the Schrieffer-Wolff mechanism. It is described by \cite{fulde:85}
\be
H_{ex}
\!
=
\!
-\sum_i\int d\br I_{ex}(\br-\bR_i)\bS_i\cdot\bsig(\br)=-I_{ex}\sum_\bq\bS_\bq\bs_{-\bq},
\label{eq:hex}
\ee
where we introduced the Fourier components with momentum \bq~for  localised and conduction spins in the second equation.
We adopt here the convention that onsite exchange $I_{ex}$ and also effective intersite exchange $J$ below  are positive or negative for FM or AF cases, respectively.
 In the  second order perturbation theory with respect to the {\it onsite} exchange $I_{ex}$ one then obtains in the conventional way the planar (2D) effective {\it intersite} RKKY local moment exchange $J(\bq)=-I^2_{ex}\chi_0(\bq)$ where \bq~ lies within the 2D Brillouin zone (BZ) of a square lattice. Here the static conduction electron susceptibility is obtained from
\be
\chi_0(\bq)=\sum_\bk\frac{f_{\bk+\bq}-f_\bk}{\eps_{\bk}-\eps_{\bk+\bq}},
\label{eq:sus0}
\ee
where $f_{\bk}$ denotes the Fermi-Dirac distribution function. We use this expression in the numerical calculation, therefore the  {\it on-site} $I_{ex}$ (or equivalently $\tga=SI_{ex}$) is to be considered the only exchange parameter in the model.  Its size can be estimated in the rare earth series by considering the quasielastic linewidth of dilute rare earth ions in a conducting host \cite{fulde:85} which allows to extract the product $\rho_{ex}=N_0I_{ex}$ where $N_0$ is the conduction electron density of states (DOS) at the Fermi level $(\epsilon_F=0)$ (Fig.\ref{fig:disp-DOS}(a)). Typical values for $\tga$  are discussed in Sec.~\ref{sec:numerical}.
Furthermore we introduce an additional uniaxial anisotropy parameter $\Delta$ assuming the intersite exchange is of the uniaxial $xxz$-type.  Here $z$ denotes the coordinate perpendicular to the surface which is also the direction of the ordered moment. Then
\be
H_{int}=-\fs\sum_\bq J(\bq)[S^x_\bq S^x_{-\bq}+S^y_\bq S^y_{-\bq}+\De S^z_\bq S^z_{-\bq}],
\ee
with $\Delta$ denoting the anisotropy parameter. This means $\Delta >1$ with the ordered moments  $\la S_\bQ^z\ra$ aligned with the $z$ axis. 
 In the 4f materials the exchange anisotropy parameter $\Delta$ for in/out-of plane moment direction is fixed
by the precise composition of the crystalline electric field (CEF) ground state which determines the matrix elements of the total angular momentum operators $J_\alpha$ ($\alpha=x,y.z$)  that constitutes the pseudospin algebra of a Kramers degenerate ground state. Their effective exchange may then be nearly isotropic ($\Delta\simeq 0$) as well as Ising-type ($\Delta \gg 1$) or any value in between.


As mentioned in the introduction we restrict to a near-nesting situation for the Fermi surface such that the main maximum of $\chi_0(\bq)$ occurs at $\bQ=(\pi,\pi)$ with $J\equiv J(\bQ)<0$ corresponding to commensurate AF order of the local moments \cite{akbari:25}. Due to the onsite exchange interaction of Eq.~(\ref{eq:hex}) the staggered local  moments $\la S_\bQ^z\ra$  produce a molecular field that reconstructs the conduction electron bands. Their total Hamiltonian (kinetic term and molecular field term)  is then
given by
\be
H_c=\sum_{\bk\si}\bigl[\epsilon_\bk c^\dag_{\bk\si}c_{\bk\si}-
\fs\si\tga(c^\dag_{\bk+\bQ\si}c_{\bk\si}+c^\dag_{\bk\si}c_{\bk+\bQ\si})\bigr],
\label{eq:Hc1}
\ee
where $\epsilon(\bk)=-2t(\cos k_x+\cos k_y)-4t'\cos k_x\cos k_y$
is the tight-binding band dispersion on a two-dimensional square lattice,
with first- and second-neighbor hopping amplitudes $t$ and $t'$, respectively.
Here $\tga=SI_{ex}$ is the effective coupling strength to the localised AF
ordered moments, which also gives
$J=-(\tga/S)^2\chi_0(\bQ)$ for the AF intersite exchange constant.
This Hamiltonian may be written in matrix form convenient for the Green's function derivation below as
\be
H_c=\fs\sum\psi^\dag_{\bk\si}\hh^c_{\bk\si}\psi_{\bk\si};\;\;\ 
\hh^c_{\bk\si}=
\left(
\begin{array}{cc}
\epsilon_\bk& -\si\tga\\
-\si\tga&\epsilon_{\bk+\bQ}
\end{array}
\right),
\label{eq:Hc2}
\ee
with the 2-component spinor $\psi^\dag_{\bk\si}=(c^\dag_{\bk\si},c^\dag_{\bk+\bQ\si})$. The molecular field describes only the static effect
of AF order leading to certain reconstructions (magnetic gaps) of the conduction band which will cause the anomalous tunnelling spectral features at the AF satellite wave vectors $\pm\bQ$ to appear.  However the primary focus here is on the important issue whether the dynamics of local magnetic moments, i.e. the AF magnon excitations,  lead to an additional characteristic imprint in the tunnelling signature of conduction electrons which could be used to identify some properties of the surface magnon spectrum.\\

%
%
Now we briefly describe the derivation of magnon modes for the $xxz$ model \cite{majlis:07,binghao:02}.
Using the Holstein--Primakoff (HP) representation for the AFM on the two
sublattices A and B,
$$
S_A^z=S-a^\dag a,\qquad
S_A^+=\sqrt{2S}\,a,\qquad
S_A^-=\sqrt{2S}\,a^\dag ,
$$
and
$$
S_B^z=-S+b^\dag b,\qquad
S_B^+=\sqrt{2S}\,b^\dag,\qquad
S_B^-=\sqrt{2S}\,b ,
$$
followed by the Fourier transformation
$$
a^\dag_\bq=\left(\frac{N}{2}\right)^{-\frac12}
\sum_{j\in A} e^{-i\bq\cdot\bR_j}a_j^\dag ,
$$
with an analogous expression for the B sublattice, we obtain the bilinear
Hamiltonian
\be
\bl
&H_{int}=-\frac{N}{2}z\De|J|S^2
\\
&+{ z|J|}S\sum_\bq\bigl[
\De(a_\bq^\dag a_\bq+b_\bq^\dag b_\bq)+\gamma_\bq(a_\bq b_{-\bq} + a^\dag_{\bq}b^\dag_{-\bq})\bigr],
\label{eq:Hbilin}
\el
\ee
where $z$ is the coordination number of bonds to N.N sites $\bdel$  and $\gamma_\bq=z^{-1}\sum_{\bdel}\exp(i\bq\bdel)$ the corresponding
structure function. For the 2D square surface lattice case (lattice constant $a=1$) to which we restrict we have $\gamma_\bq=\fs(\cos q_x+\cos q_y)$.
The summation in Eq.~(\ref{eq:Hbilin}) runs over the 2D reduced AF Brillouin zone (AFBZ) defined by $|q_x+q_y|\leq\pi$ (Fig.~\ref{fig:FS-spec}(a)).

The bilinear AF magnon Hamiltonian may be diagonalised with a Bogoliubov transformation defined by $a_\bq=u_\bq\al_\bq+v_\bq\beta^\dag_{-\bq}$ and
$b_\bq=u_\bq\beta_\bq+v_\bq\al^\dag_{-\bq}$ . The coefficients can be parameterised as $u_\bq=\cosh\Phi_\bq$ and $v_\bq=\sinh\Phi_\bq$ since they
must satisfy $u_\bq^2-v_\bq^2=1$ to fulfil the bosonic commutation relations for the magnon operators $\al_\bq,\beta_\bq$. The condition for diagonalising the above Hamiltonian is then simply
\be
\tanh 2\Phi_\bq=-\frac{\gamma_\bq}{\De},
\ee
which leads to the free magnon Hamiltonian of the AFM ordered state according to 
\be
\bl
H_{int}=&E_0+\sum_\bq\omega_\bq(\al^\dag_\bq\al_\bq+\beta^\dag_\bq\beta_\bq)=E_0+H_{m}.
\label{eq:Hmag}
\el
\ee
Here 
$$E_0=-\frac{N}{2}z\De|J|S^2-\frac{N}{2}z\De|J|S+\sum_\bq\omega_\bq$$
 is the ground state energy of the $xxz$ model AFM state  and $H_m$ the magnon Hamiltonian with  $\omega_\bq$ denoting the dispersion of the two {\it degenerate} magnon modes created by $(\alpha^\dag_\bq,\beta^\dag_\bq)$. It is simply given by
\be
\omega_\bq=z|J|S(\De^2-\gamma^2_\bq)^\fs,
\label{eq:magdisp}
\ee
which is periodic with the new AFBZ reciprocal lattice vector $\bQ$ according to $\omega_{\bq+\bQ}=\omega_\bq$.
With the easy axis anisotropy $\De>1$ the spectrum is gapped and becomes gapless in
in the isotropic Heisenberg limit $\De=1$ with a linear magnon dispersion $\omega_\bq= v_m|\bq|$ for small momentum 
and a corresponding magnon velocity $v_m=(1/\sqrt{2})z|J|S$. The extremal values are the minimum at $\bq =(0,0),\bQ=(\pi,\pi)$ ($\Gamma$, M- points) with $\omega_{\Gamma,\text{M}}=\omega_0(\De^2-1)^\fs$ ($\omega_0:=z|J|S$) and the maximum at $\bq=(\pi,0),(0,\pi)$ (X- point) with  $\omega_X=\omega_0\De$. The latter corresponds to a van Hove singularity leading to the magnon DOS peak at this energy  (see Fig.~\ref{fig:magnon-disp}). The resulting magnon dispersion width given by $D_m/\omega_0=\Delta-(\Delta^2-1)^\fs$ with the limiting cases  $D_m/\omega_0\simeq 1-(2\delta)^\fs$ for  nearly isotropic $\Delta=1+\delta;\; \delta\ll 1$ and $D_m/\omega_0\simeq 1/(2\Delta)$ for  near Ising- case $\Delta \gg 1$.  The ratio of maximum to minimum magnon energies is  $\kappa_m=\Delta/(\De^2-1)^\fs$, leading to $\kappa_m=
2.4$ in the mostly used case of $\De=1.1$ (Fig.~\ref{fig:magnon-disp}(a)). Such small values corresponding to a large magnon gap are entirely possible in 4f materials  \cite{rotter:12} due to wide range of possible intersite exchange anisotropies introduced in Sec.~\ref{sec:model}.

With the help of HP and Bogoliubov transformation to magnon
coordinates, we can now also derive the dynamical interaction of conduction electrons with the latter, in addition to the static
AFM staggered molecular field effect contained in $H_c$ (Eqs.~(\ref{eq:Hc1},\ref{eq:Hc2})). Introducing the conduction
electron spin components $s^+_\bq=\fs\sum_\bk c^\dag_{\bk-\bq\ua}c_{\bk\da}$ and  $s^-_\bq=\fs\sum_\bk c^\dag_{\bk-\bq\da}c_{\bk\ua}$ and expressing the localised spin components $S_\bq^\pm$ by the magnon operators the dynamical part of $H_{ex}$ is now obtained from the original form Eq.~(\ref{eq:hex}) as
\be
\bl
H_{cm}
=
&
-I_0\sum_{\bk\bq}(u_\bq+v_\bq)
\times\\
&\bigl[c^\dag_{\bk-\bq\da}c_{\bk\ua}(\al_\bq+\beta^\dag_{-\bq})
 +c^\dag_{\bk-\bq\ua}c_{\bk\da}(\al^\dag_{-\bq}+\beta_\bq)\bigr],
 \label{eq:Hcm}
\el
\ee
where we used the identity 
$a_\bq +b^\dag_{-\bq}=(u_\bq+v_\bq)(\al_{\bq} +\beta^\dag_{-\bq})$.
and defined the effective conduction electron-magnon coupling constant by $I_0=\fs\sqrt{\frac{S}{2}}I_{ex}$. The summation over magnon wave vectors $\bq$ again runs over the reduced AFBZ whereas the summation over conduction electron momentum \bk~covers the full 2D square lattice Brillouin zone (BZ) with $|k_x|,|k_y|\leq\pi$. The total Hamiltonian of the dual conduction electron - $xxz$ localised spin model is then finally obtained
as the sum $H=H_c+H_m+H_{cm}$ from Eqs.~(\ref{eq:Hc1},\ref{eq:Hmag},\ref{eq:Hcm}), respectively.
%
\begin{figure}
\includegraphics[width=\columnwidth]{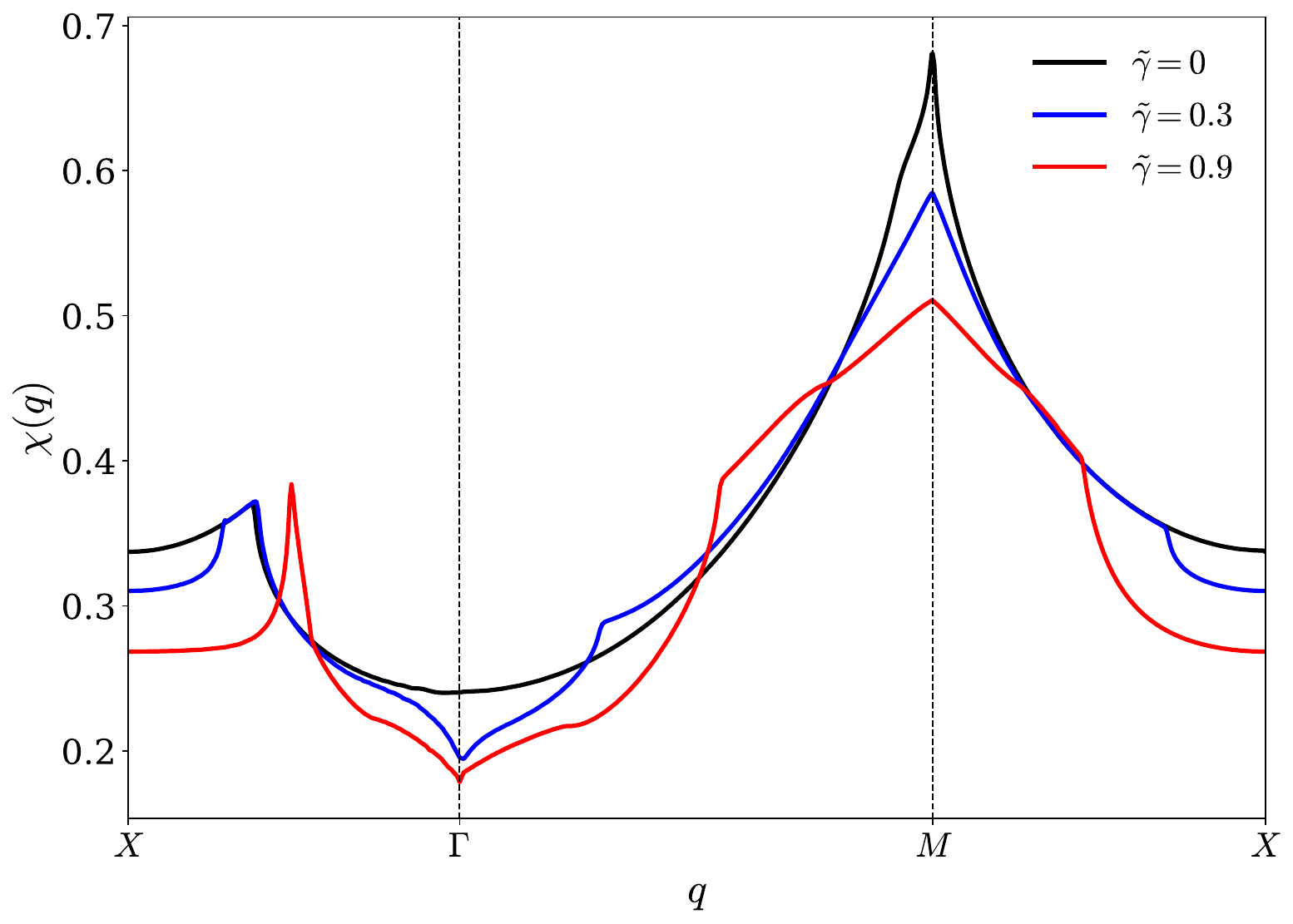}
\caption{Static susceptibility $\chi(\bq)$ for bare bands ($\tga=0$, black line) in the paramagnetic state 
with the maximum located at the M point corresponding to ordering
vector \bQ=($\pi,\pi$)  in units of $1/t$. In the AF ordered case with reconstructed  conduction bands 
by nonzero $\tga$ the maximum rests at $\bQ$ but is progressively diminished due to the AF gap opening
and corresponding suppressed nesting property (see Fig.~\ref{fig:FS-spec}).   In this and all following BZ path
plots the $\Gamma X$ sector has a length $\pi/a$ with $a\equiv 1$ and the sectors are scaled accordingly.}
\label{fig:sus}
\end{figure}
%

\section{The tunnelling spectrum from the magnon renormalised Green's functions}
\label{sec:current}

The central issue of this work is to identify to which extent the spectrum of magnons in an AFM ( or general helical) local moment  magnet have an influence on the STM tunnelling conductance of the conduction electrons which interact with the magnons as described by $H_{cm}$. The tunnelling surface conductance which is measured as function of the tip position $\br$ is obtained in a simple Bardeen-type approach \cite{hoffman:11} from
\be
\frac{dI}{dV}(\br)\approx -\frac{4\pi e^2}{\hbar}|M|^2\rho_{tip}(0)\rho(\omega=eV,\br).
\label{eq:STM}
\ee
Here I(V) is the total current and V the applied bias voltage to the metallic  (but nonmagnetic) tip.
Furthermore $\rho_{tip}(0)$ is the electronic DOS and $|M|^2$ the square of the tunnelling matrix element.  It is assumed that both are slowly varying with electron energy $\omega=eV$, i.e., the applied voltage. Then the scan of the  tunnelling conductance determines the  spatially dependent conduction electron  DOS $\rho(\omega,\br)$  at the surface of the sample. It has contributions from the 
background DOS of the lattice at the surface and from modulations coming from i) reconstruction of conduction states by the static localised AF molecular field and ii) the scattering from magnons of the localised moments resulting in a dynamic self energy of conduction electrons. For helical magnetic order it was shown experimentally in Ref.~\onlinecite{spethmann:24} and theoretically in Ref.~\onlinecite{akbari:25} that the former leads to additional Bragg-type satellite peaks at the magnetic ordering wave vector, even though conduction electrons are not the primary medium of magnetic order. This was attributed to the molecular field part of $H_c$ created by the underlying local moment order \cite{akbari:25}. The second dynamic influence of itinerant electron scattering from local moment magnons will now be included on the same footing. In this work we do not discuss the simultaneous influence of surface impurity scattering leading to the quasiparticle interference phenomenon \cite{akbari:25}. To be able to isolate the effect of scattering from magnons we restrict to the pure surface model.

\subsection{Real space Green's functions and satellite peaks}
\label{sec:spectrum}

For this purpose we are using the real space Green's function representation for the conduction electrons of the pure surface 
as defined by
\be
G_0(\br,\br',\si,\om)=\sum_\bk\Psi_{\bk\si}^\dag(\br')[\om-\hh_\bk]^{-1}\Psi_{\bk\si}(\br),
\label{eq:Green1}
\ee
where the conduction states in spinor presentation are given by $\Psi_{\bk\si}^\dag(\br')=(\psi^*_{\bk\sigma}(\br'),\psi^*_{\bk+\bQ\sigma}(\br'))$.
In plane wave approximation we have $\psi_{\bk\si}(\br)=(1/\sqrt{N}){\rm  exp}(i\bk\cdot\br)\chi_\si$ where $\chi_\si$ is the two-component  spin wave function (for general Bloch states see the discussion in Ref.~\onlinecite{akbari:25}). The latter are not mixed by the molecular field since we have an easy axis AF state.
The Fourier transform of the matrix Green's function in the two-component spinor space is obtained from Dyson's equation 
$$\hG(\bk\si,\om)^{-1}=\hG_0(\bk\si,\om)^{-1}-\hat{\Sigma}(\bk\si,\om)$$
 with $\hat{\Sigma}$ denoting the self energy matrix due to $H_{cm}$. This  can be inverted to give
\be
\bl
&\hG(\bk\si,\om)=
D_\bk(\om)^{-1}
\times
\\
&\left(
\begin{array}{cc}
\!
\om-\epsilon_{\bk+\bQ}-\Sigma(\bk+\bQ,\om)& -\si\tga\\
-\si\tga&\om-\epsilon_{\bk}-\Sigma(\bk,\om)
\end{array}
\right)
\\
&=
\left(
\begin{array}{cc}
g_a& g_c\\
g_c&g_b
\end{array}
\right).
\label{eq:Green2}
\el
\ee
Here the determinant of $\hG(\bk\si,\om)^{-1}$ is given by
\be
\bl
D_\bk(\om)
=&
\bigl[\om-\epsilon_{\bk}-\Sigma(\bk,\om)\bigr]
\times\\
&
\bigl[\om-\epsilon_{\bk+\bQ}-\Sigma(\bk+\bQ,\om)\bigr]-\tga^2.
\el
\ee
Its complex valued zeroes (or poles of $\hG(\bk\si,\om)$) may be interpreted as the renormalised
conduction bands with a finite lifetime due to magnon emission/absorption processes (see Eq.~(\ref{eq:qpbands})).
The Green's function and its pole positions contain both effects of magnetic order: i) the mixing of $|\bk~\si\ra$ and $|\bk+\bQ~\si\ra$ conduction states in the nondiagonal elements $\sim\tga$ and ii) the coupling of conduction electrons to virtual magnon excitations via the diagonal (spin-independent) self energies $\Sigma(\bk,\om)$ and
$\Sigma(\bk+\bQ,\om)$ to be considered in detail in the next section. Both effects of AFM local moment order are treated here on the same footing and will both modify the dispersion and lifetime of conduction states and henceforth the tunnelling current via their influence on the $\bk$-resolved conduction electron DOS. The latter is obtained from
\be
\rho_\si(\omega,\br)=-\frac{1}{\pi}\text{Im}G(\br,\br,\omega+i\eta)],
\ee
with the spatial representation of the total Green's function given in terms of the matrix elements in Eq.~(\ref{eq:Green2})  
\be
\bl
G(\br,\br,\si,\omega)=
&
\frac{1}{N}\sum_\bk\{ 
[g_a(\bk,\omega)
\!+\!
g_b(\bk,\omega)]\\
&+
g_c(\bk,\si,\omega)(e^{i\bQ\cdot\br}
\!+\!
e^{-i\bQ\cdot\br})\},
\label{eq:purelocal}
\el
\ee
which leads to the Fourier transformed spectral functions
\be
\rho_\si(\bq,\omega)=\rho_0(\omega)\delta_{\bq, 0}
+\rho^\si_{AF}(\omega)[\delta_{\bq +\bQ}+\delta_{\bq -\bQ}],
\label{eq:totDOS1}
\ee

%
\begin{figure}
\includegraphics[width=\columnwidth]{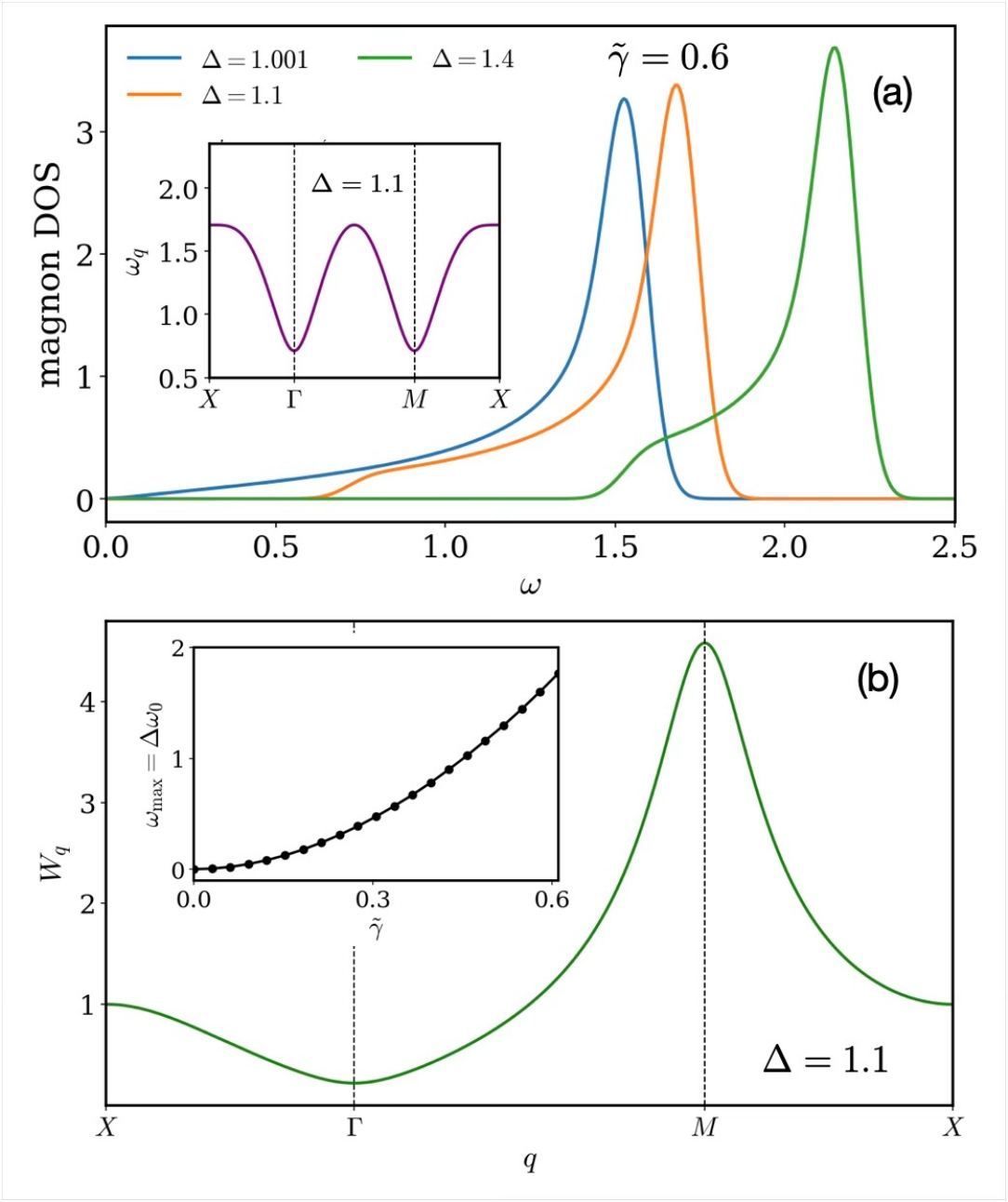}
\caption{(a) Inset shows the magnon dispersion in the original BZ for
$\tga=0.6$, $\De=1.1$ leading to  $|J|=I_{ex}^2\chi(\bQ)\approx 0.77$, $\omega_0=1.54$ and $\omega_{max}=\De\omega_0=1.7$. Main figure
 presents corresponding magnon DOS shifting progressively to higher energy with increasing $\Delta$.
(b) Coherence factor of electron-magnon interaction. It strongly peaks at the M-point
where the magnon dispersion  has its minimum with corresponding  low DOS at this energy $\omega=\omega_0(\De^2-1)^\fs$.
Note that the dispersion is periodic with $\omega_{\bq+\bQ}=\omega_\bq$ while the coherence factor 
is not since $W_{\bq+\bQ}=1/W_\bq$.}
\label{fig:magnon-disp}
\end{figure}
%

The first part is the  normal homogeneous background spectral function  $\rho_0$  due to assumed plane wave conduction  states and it corresponds to a peak at $\bq=0$. Its $\bk$-space integral is the conduction electron DOS. The second anomalous part $\rho_{AF}$  is due to the reconstruction of conduction bands (leading to magnetic subgap openings)  and mixing of $|\bk\ra$ and $|\bk+\bQ\ra$ states caused by the static AF molecular field terms as discussed in detail in Ref.\cite{akbari:25} for the incommensurate helical case.  The mixing leads to nondiagonal elements in the Green's function
matrix of Eq.(\ref{eq:Green2}) which contribute in the real space representation Eqs.(\ref{eq:purelocal}) as spacially modulated parts. Their most important aspect is the corresponding appearance of anomalous spectral {\it satellites} at the AFM ordering wave vector $\bq=\pm\bQ$ associated with $\rho^\sigma_{AF}$ in Eq.~(\ref{eq:totDOS1}).
 Although the above equation is formally the same as in the static case both normal and anomalous spectral functions now contain the modifications caused by the {\it dynamical  magnon self energy} which will lead to characteristic changes, in particular at the 
additional AF induced satellite positions $\pm\bQ$. 
Explicitly, after analytic continuation $\om\rightarrow \omega+i\eta$, we have
\be
\bl
&
\rho_0(\omega)
=
\frac{-1}{\pi N}
{\rm Im}
\sum_\bk \frac
{2[\om-\fs(\tep_{\bk+\bQ}+{ \tep_\bk)}]}
{(\om-\tep_{\bk+\bQ})(\om-\tep_{\bk})-\tga^2},
\\
&
\rho^\si_{AF}(\omega)
=
\frac{\sigma}{\pi N}
{\rm Im}
\sum_\bk 
\frac
{\tga}
{(\om-\tep_{\bk+\bQ})(\om-\tep_{\bk})-\tga^2},
\el
\ee
with
\be
\bl
&
\tep_\bk(\om)
=
\eps_\bk+\Sigma(\bk,\om),
\\
&
\tep_{\bk+\bQ}(\om)
=
\eps_{\bk+\bQ}+\Sigma(\bk+\bQ,\om),
\label{eq:STMspectra}
\el
\ee
where $\tep_\bk$ may be interpreted as {\it unreconstructed} but magnon self energy modified conduction bands.
For clarity we note that for general conduction electron Bloch
functions $\rho_0(\bq,\om)$ will also have peaks appearing at the reciprocal lattice vectors $\bK$ of the
full (nonmagnetic) BZ and likewise the anomalous part will develop AFM satellites $\bK\pm\bQ$ around all $\bK$ positions~\cite{akbari:25}.\\

%
\begin{figure}
\includegraphics[width=0.99\columnwidth]{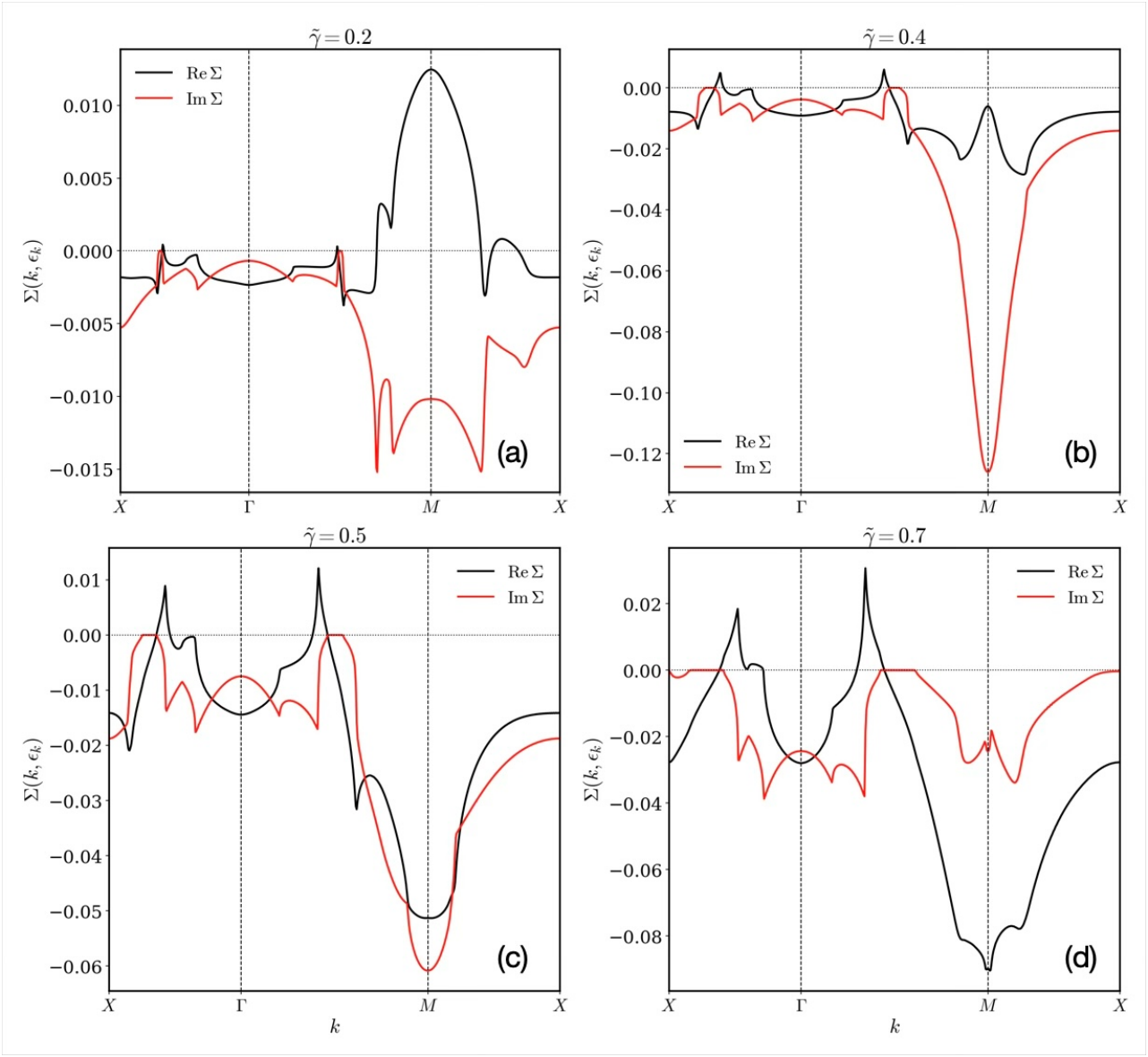}
\caption{ Conduction electron self energies $\Sigma(\bk,\epsilon_\bk)$ along X$\Gamma$MX BZ path 
(real and imaginary parts) caused by scattering from magnons according
to Eq.~(\ref{eq:selfnum}) for weak, intermediate and strong 
exchange couplings. As $\tga$ is increased
a shift of weight from the M-point to the AF equivalent $\Gamma$ -point is observed for the imaginary part and the real
part at the M point changes sign.}
\label{fig:self-disp}
\end{figure}
%

The expressions for the quasiparticle bands are obtained from the zeroes of  $D_\bk(\om)$.
When we {\it exclude} the self energy due to magnon coupling we obtain the conduction bands reconstructed only by the static AF order as
\be
E^\pm_\bk=\fs(\eps_{\bk+\bQ}+\eps_\bk)
\mp\{\frac{1}{4}(\eps_{\bk+\bQ}-\eps_\bk)^2+\tga^2\}^\fs.
\label{eq:recon-disp}
\ee
Here (different from the previous convention \cite{akbari:25}) we used the band ordering according to energies leading to band interchange at the crossing points. 
The susceptibility $\chi(\bq)$ for this case shown by the blue and red curve in Fig.~\ref{fig:sus} is obtained by replacing $\eps_\bk\rightarrow E^\pm_\bk$ in Eq.~(\ref{eq:sus0}) for $\chi_0(\bq)$ ($\tga=0$; black curve)  and summing over both bands. Likewise the spectral functions in Eq.~(\ref{eq:STMspectra}) in that limit reduce to \cite{akbari:25}
\be
\bl
\rho^{(0)}_0(\bk,\omega)
=
&\sum_{n=\pm}\delta(\omega-E_\bk^n),
\\
\rho_{AF}^{(0)\si}(\bk\omega)
=
&\si\sin\theta_\bk\cos\theta_\bk\sum_{n=\pm}n\delta(\omega-E_\bk^n),
\label{eq:DOSstatic}
\el
\ee
where we used  ${\tga/(E_\bk^--E_\bk^+)}$$=sin\theta_\bk\cos\theta_\bk$.
Here the mixing angle $\theta_\bk$ of $|\bk\ra$ and $|\bk+\bQ\ra$ states controls the reconstruction of conduction bands by the static AF order parameter and is given by
\be
\tan 2\theta_\bk=\frac{\tga}{d_\bk};\;\;d_\bk=\fs(\eps_{\bk+\bQ}-\eps_\bk),
\ee
Note that the frequency integral over the anomalous spectrum $\rho_{AF}^\si(\bk,\omega)$ vanishes because the exchange interaction with the background AF order cannot establish a charge density wave with the same wave vector \bQ.\\

The reconstructed quasiparticle bands {\it including} now the magnon interactions are approximately obtained by
replacing $\Sigma(\bk,\om)\rightarrow \Sigma(\bk,E^\pm_{\bk})$ in Eq.~(\ref{eq:Green2}), which leads to the expressions
\be
\bl
\tE^\pm_\bk
=&
\fs(\eps_{\bk+\bQ}+\eps_\bk)+\Sigma_s(\bk,E^\pm_\bk)
\\
&\mp\{\bigl[\fs(\eps_{\bk+\bQ}-\eps_\bk)+\Sigma_a(\bk,E^\pm_\bk)\bigr]^2+\tga^2\}^\fs,
\label{eq:qpbands}
\el
\ee
with the (anti-)symmetric self energies defined by $\Sigma_{a,s}(\bk,\omega)=\fs[\Sigma(\bk+\bQ,\omega)\mp\Sigma(\bk,\omega)]$, respectively.  The real part of $\tE^\pm_\bk$ gives the reconstructed quasiparticle dispersion including renormalization by magnons while the modulus of the imaginary part corresponds to their inverse lifetime $1/\tau^\pm_\bk$. The change of the electronic dispersion originating from the self energy (Fig.~\ref{fig:self-disp}) as compared to the static reconstructed dispersion in Eq.(\ref{eq:recon-disp}) is given
by $\delta \tE^\pm(\bk)=\tE^\pm(\bk)-E^\pm(\bk)$ and shown in Fig.~\ref{fig:quasidisp} together with the broadening.
The latter is most prominent at the $\Gamma$, M -points of the full BZ  where the magnon energy is small 
and it vanishes at the $X'(\frac{\pi}{2},\frac{\pi}{2})$ AFBZ zone boundary points where the AF gap opens.\\

In terms of these approximate complex valued dispersions the spectral functions in Eq.~(\ref{eq:STMspectra}) may be rewritten in
shorthand notation as
\be
\bl
\rho_0(\bk,\omega)
=&
-\frac{1}{\pi}
{\rm Im}
\sum_{n=\pm}
(\omega-\tE^n_\bk +i\eta)^{-1}
\\
\rho_{AF}(\bk,\omega)
=
&
-\frac{\si}{\pi}
{\rm Im}
\sum_{n=\pm}n\tga 
\Big[
\frac{1}{\tE^+_\bk-\tE^-_\bk}
\Big]
\Big[
 \frac{1}{\omega-\tE^n_\bk +i\eta}
 \Big]
.
\label{eq:fullspecshort}
\el
\ee
Thus $\rho_0(\bk,\omega)$ can be seen again to represent the normal spectral functions of quasiparticle bands while the anomalous $\rho_{AF}(\bk,\omega)$ appears only due to the coupling $\tga$ to the local moment background AF order. These expressions reduce to Eq.~(\ref{eq:DOSstatic}) when the complex valued $\tE_\bk^n$ dispersions including the magnon
self energy are replaced by the real valued $E^n_\bk$ of Eq.~(\ref{eq:recon-disp}) including only the static reconstruction.
From the form of Eq.~(\ref{eq:fullspecshort}) it is obvious that the frequency integral of $\rho_0(\bk,\omega)$ is the total occupation $n_0=2$
of the band and that of  $\rho_{AF}(\bk,\omega)$ is zero; as in the purely static case (Eq.~(\ref{eq:DOSstatic})) without magnon self energy.
For numerical calculations it is, however, preferable to use the original expressions in Eq.~(\ref{eq:STMspectra}) for the  spectral functions.
%
\begin{center}
\begin{table*}
\caption{Localised-itinerant model parameters and their relations. The energy scale is set
by n.n. hopping $t\equiv 1$. As independent model parameters we then
have  $t'$; $\Delta$; $\tga$, and the Fermi energy $\epsilon_F$. In all figures we fix
$t'=0.4$ and $\epsilon_F=0$, leaving only the size of the exchange coupling
$\tga$ and the exchange anisotropy $\De$ as variable parameters.}
\vspace{0.5cm}
\begin{center}
\begin{tabular}{c @{\hspace{4mm}} c @{\hspace{4mm}} c }
\hline\hline
parameter & designation &\\
\hline
TB electronic hopping elements & $t, t'$ & \\
localised (pseudo-)spin size & $S (=\fs)$&\\
on-site cf- exchange& $I_{ex}$ & \\
inter-site ff- exchange& $J=-I^2_{ex}\chi_0(\bQ)<0$&  \\
ff-exchange anisotropy& $\Delta$&\\
band reconstruction parameter & $\tga=SI_{ex}$&\\
magnon energy scale & $\omega_0=z|J|S$ $(z=4)$ &\\
electron magnon interaction    & $I_0=\frac{\tga}{2\sqrt{2S}}=\fs\sqrt{\frac{S}{2}}I_{ex}$ &   \\
\hline\hline
\end{tabular}
\end{center}
\label{tbl:matel}
\end{table*}
\end{center}
%

\subsection{Electron self energy due to interaction with magnons }
\label{sec:selfenergy}

\begin{figure}
\includegraphics[width=\columnwidth]{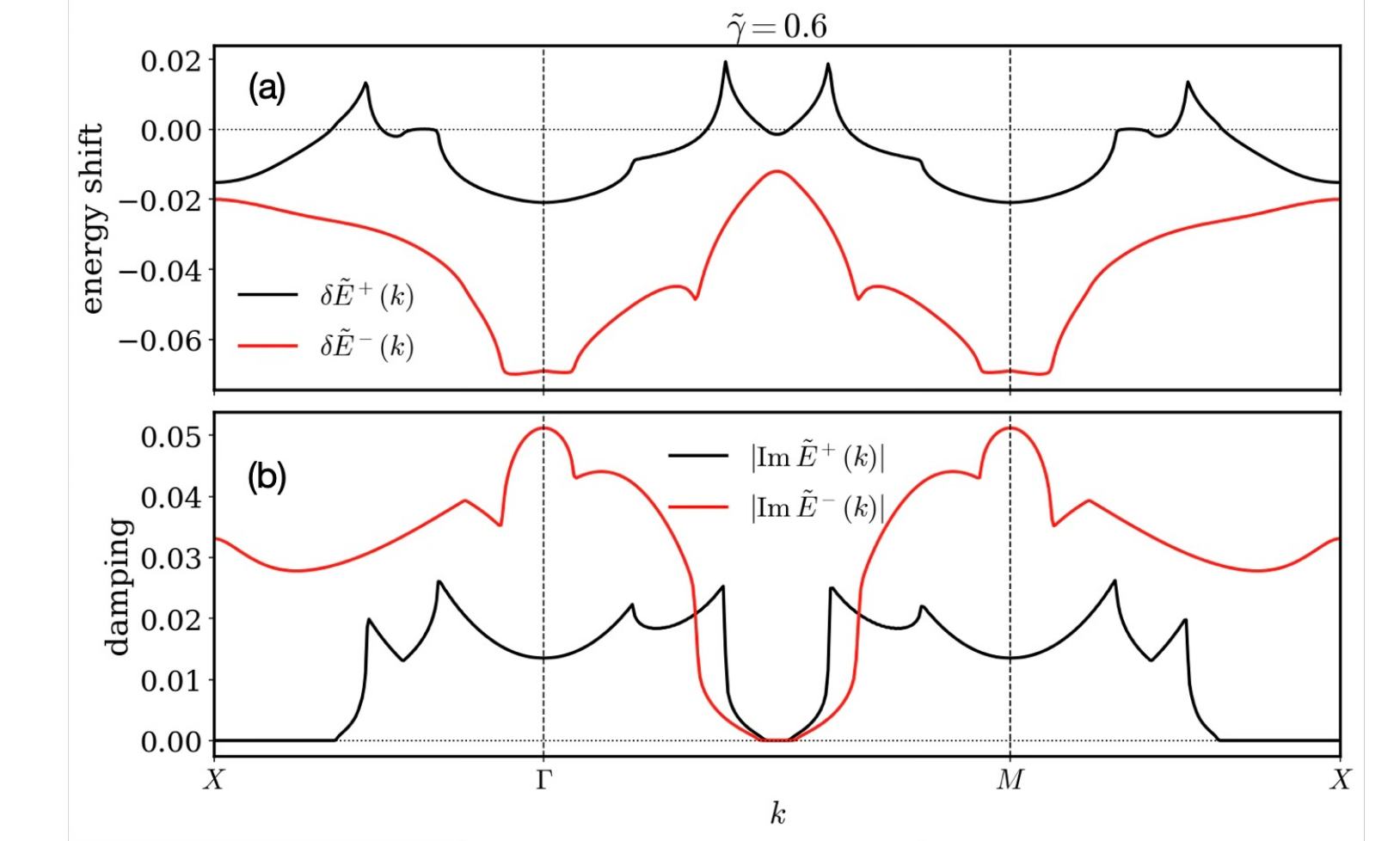}
\caption{Quasiparticle renormalisation due to coupling to magnons (Eq.~(\ref{eq:qpbands})). (a) Shift of band energies $\delta \tE^\pm=\tE^\pm-E^\pm$. (b) Imaginary part of quasiparticle energy leading to the broadening $\Gamma^\pm= |{\rm Im} E^\pm|$ of dispersion. It vanishes at the $X'(\frac{\pi}{2},\frac{\pi}{2})$ AFBZ zone boundary point where the AF gap opens.}
\label{fig:quasidisp}
\end{figure}

According to previous section the central task to determine the influence of dynamical magnon coupling in the STM spectrum  is the determination of the electron self energy $\Sigma(\bk,\om)$, similar as in Refs.\onlinecite{maeland:21,revenda:25,srivastava:05,richmond:70}. The self energy will be restricted to contributions from virtual one-magnon emission/absorption processes. In this approximation, using 
the electron-magnon interaction of Eq.~(\ref{eq:Hcm}) we obtain the self energy to second order in $I_0$ as
\be
\bl
&
\Sigma(\bk,\om,\si)
=
\\
&
\!-
\!
\frac{I_0^2}{N}\sum_\bq(u_\bq+v_\bq)^2
\!
\sum_\lam T
\sum_{\omega_\nu}D_\bq^{0\lam}(i\omega_\nu)G_{\bk+\bq,\bar{\si}}(\om
\!
+
\!
i\omega_\nu).
\label{eq:self1}
\el
\ee
Here $D_\bq^{0\lam}(i\omega_\nu)=\lam (i\omega_\nu-\lam\omega_\bq)^{-1};\; (\lam=\pm 1)$ are the magnon and $G_{\bk\,\om,\sigma}=(\om-\eps_\bk)^{-1}$ the electron propagators, respectively. The spin $\sigma=-\bar{\sigma}$ appears only as dummy variable, furthermore $\omega_n=(2n+1)T$ and $\omega_\nu=2\pi\nu T$ are Matsubara frequencies.
This has the same form as an electron-phonon self energy, except for the coherence factors due
to the underlying AFM structure. They can be evaluated as
\be
W_\bq=(u_\bq+v_\bq)^2=\Bigl(\frac{\De-\gamma_\bq}{\De+\gamma_\bq}\Bigr)^\fs,
\label{eq:cohfac}
\ee
which is always finite for $\De >1$. At the symmetry points we have the values $W_\Gamma=\Bigl(\frac{\De-1}{\De+1}\Bigr)^\fs, W_X=1,W_M=\Bigl(\frac{\De+1}{\De-1}\Bigr)^\fs$, for $\De=1.1$ used in most calculations their numerical values are 0.22, 1, 4.58, respectively, independent of $\tga$ (cf. Fig.~\ref{fig:magnon-disp}(b)). 
We notice that $W_{\bq+\bQ}=W_\bq^{-1}$. Therefore the self energies $\Sigma(\bk+\bQ,\om)$ and $\Sigma(\bk,\om)$ appearing in the spectra of Eq.~(\ref{eq:STMspectra}) are not equivalent despite $\omega_{\bq+\bQ}=\omega_\bq$.
However the renormalized quasiparticle bands in Eq.~(\ref{eq:qpbands}) again exhibit the required periodicity 
$\tE^\pm_{\bk+\bQ}=\tE^\pm_\bk$ in the AFBZ (cf. Fig.~\ref{fig:quasidisp}).
Carrying out the Matsubara summation in  Eq.~(\ref{eq:self1}) at taking the limit $T\rightarrow 0$
then leads to the explicit self energy expression
\be
\bl
&
\Sigma(\bk,\om,\si)
=
\frac{I_0^2}{N}\sum_\bq 
\Bigl(\frac{\De-\gamma_\bq}{\De+\gamma_\bq}\Bigr)^\fs
\times
\\
&\qquad
\Bigl[
\frac{f_0(\eps_{\bk+\bq})}{\omega-\eps_{\bk+\bq}+\omega_\bq+i\delta}
+\frac{1-f_0(\eps_{\bk+\bq})}{\omega-\eps_{\bk+\bq}-\omega_\bq+i\delta}
\Bigr].
\label{eq:selfnum}
\el
\ee
Here $f_0$ is the Fermi function at zero temperature, i.e. $f_0(\eps)=\Theta_H(-\hep)=1-\Theta_H(\hep)$ where $\Theta_H$ is the Heaviside function and $\hep=\eps-\eps_F$ is the energy with respect to Fermi level $\eps_F$. We stress that the $\bq$- summation extends over the AFBZ (dashed blue line in Fig.~\ref{fig:FS-spec}(a).
Inserting Eq.~(\ref{eq:selfnum}) into Eq.~(\ref{eq:STMspectra}) leads to the  regular tunnelling spectrum at $\bq=0$ and 
the anomalous AF induced spectrum at $\bq=\pm\bQ$, both simultaneously modified by band reconstruction
due to AF order and magnon scattering.\\
%
%
%
\begin{figure*}
\includegraphics[width=1.5\columnwidth]{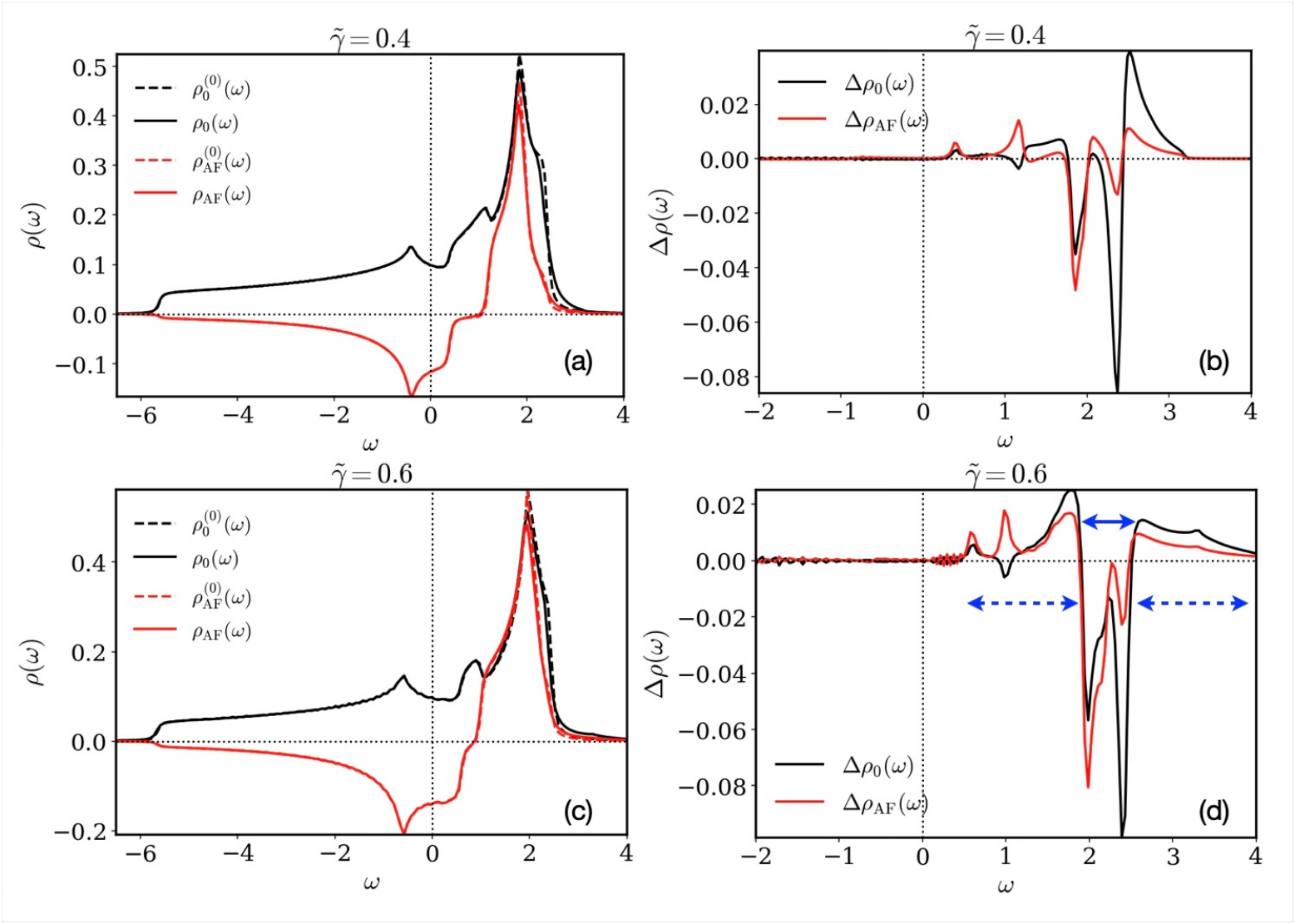}
\caption{Normal (a,c) and anomalous  (b,d) spectral functions $(\rho_{AF}=\rho^-_{AF}-\rho^+_{AF})$ according 
to Eq.~(\ref{eq:STMspectra}) for $\tga=0.4, 0.6$
for the case without (dashed lines) and with (full lines) self energy $\Sigma$  included. The latter describes the effect of 
static band reconstruction as well as the dynamic effect due to magnon scattering.
Dynamic self energy effects with magnon scattering are shown separately in (c,d) with $\Delta\rho_0=\rho_0-\rho^{(0)}_0$ 
and $\Delta\rho_{AF}=\rho_{AF}-\rho^{(0)}_{AF}$. Here upper index $(0)$ means setting $\Sigma=0$ (equivalent to Eq.~(\ref{eq:DOSstatic})). The magnon self energy shifts spectral weight from the main peak (full blue arrow) to magnon sidebands (dashed blue arrows). }
\label{fig:fullspec}
\end{figure*}
%

It is also instructive to plot the real and imaginary parts of the self energy along a path in the BZ like the electron and magnon dispersions.
For that purpose we replace $\om \rightarrow \epsilon_\bk$ in the bottom line of Eq.~(\ref{eq:STMspectra})
as a first iterative approximation. Then we obtain for the imaginary part
\be
\bl
Im\Sigma(\bk,\epsilon_\bk,\si)
=
&
-\pi\frac{I_0^2}{N}\sum_\bq 
\Bigl(\frac{\De-\gamma_\bq}{\De+\gamma_\bq}\Bigr)^\fs
\times
\\
&\Bigl[
f_0(\eps_{\bk+\bq})\delta(\eps_\bk-\eps_{\bk+\bq}+\omega_\bq)
\\
&+(1-f_0(\eps_{\bk+\bq}))\delta(\eps_\bk-\eps_{\bk+\bq}-\omega_\bq)
\Bigr],
\el
\ee
and a corresponding expression holds for the real part Re[$\Sigma(\bk,\epsilon_\bk,\si)$]. Their typical behaviour 
on the X$\Gamma$MX BZ- path may then be calculated performing the $\bq$ integration numerically (Fig.~\ref{fig:self-disp}).

\section{Numerical results and discussion: fingerprint of magnons in the STM spectrum}
\label{sec:numerical}

In this section we discuss typical numerical results that illustrate the effect of electron-magnon scattering on
the normal $(\bq=0)$ and anomalous $(\bq=\pm\bQ)$ STM spectra as given in Eqs.~(\ref{eq:totDOS1},\ref{eq:STMspectra}).
The relevant model parameters employed and their inter-relations are defined in Table~\ref{tbl:matel} and the set of independent parameters is defined in the caption. We use hopping parameters close to the nesting condition and sufficiently large $\tga=SI_{ex}$ to have sizeable AF gap appearance at the FS sections connected by the nesting vector $\bQ=(\pi,\pi)$.  For the localised pseudo-spin we take $S=\fs$ as the most common case of a $4f$ Kramers doublet CEF ground state.
 As mentioned  in Sec.~\ref{sec:model} the size of the product $\rho_{ex}=N_0I_{ex}$ may be obtained from INS experiments \cite{fulde:85}. It varies within the rare earth series due to different f-shell extension and connected hybridisation strength. For Ce  one obtains $\rho_{ex}\approx 0.1$ while for Gd  $\rho_{ex}\approx 0.03$. With $N_0\approx1/8t=1/W$ where the bandwidth W is assumed to be of order $W\approx 2$ eV the hopping energy scale is $t=0.25$ eV. With pseudospin $S=1/2$ for Ce Kramers doublet CEF ground state and real spin $S=5/2$ for Gd (no CEF splitting due to orbital L=0) we obtain, in units of t, $\tga\approx 0.4$ (Ce) and $\tga\approx 0.6$ (Gd) as estimated realistic values for the local moment-conduction electron interaction.

As a point of reference we first show the Fermi surface spectral function of the  bare conduction band (and also shifted by $\bQ$ or equivalents) in Fig.~\ref{fig:FS-spec}(a) in the full BZ with the AFBZ indicated by blue dashed line. It corresponds to the bare band dispersion shown in (b). When $\tga$ is turned on $\epsilon_\bk,\epsilon_{\bk+\bQ}$ bands are transformed to the reconstructed bands $E_\bk^\pm$ (b) and their spectral functions at the Fermi level is shown in (c,d). The flattened sphere of the original conduction band $\eps_\bk)$ is broken up into the four indented squares around the BZ X-points (AFBZ M'-points) due to the AF gap opening at the AFBZ X'-points $(\frac{\pi}{2},\frac{\pi}{2})$ at the Fermi level $\eps_F=0$ which are connected by the nesting and AF ordering vector $\bQ=(\pi,\pi)$. The gap opening is directly seen in the dispersion plot of Fig.~\ref{fig:FS-spec}(b). 

The corresponding DOS is shown in  Fig.~\ref{fig:disp-DOS} for $\tga=0.3$  as calculated from Eqs.(\ref{eq:DOSstatic}) that include only the coupling to the static AF local moment order ({\it excluding} the magnon self energy). The sharp peak of the absolute maximum in unreconstructed normal part $\rho_0(\omega)$ is due to the van Hove singularity at the X-points. For this small value of $\tga$ the reconstructed conduction band DOS (a) is only moderately affected by the X'-point gap. As shown later (Fig.~\ref{fig:fullspec})) for larger exchange coupling $\tga$ to the localised AF background the reconstructed DOS is strongly modified.  Concomitantly an anomalous spectral function $\rho_{AF}(\omega)$ appears in Eq.~(\ref{eq:totDOS1}) for finite $\tga$ at the AF ordering wave vector $\bQ$ which is shown in (b). It fulfils the sum rule of vanishing integral over frequency $\omega$ as mentioned at the end of Sec.~\ref{sec:spectrum}.

A similar behaviour under increasing exchange coupling may be identified in the static susceptibility of Fig.~\ref{fig:sus}.
In the paramagnetic phase with $\tga=0$ the nesting leads to the pronounced peak in the static susceptibility $(\chi=\chi_0)$ at the M-point vector $\bQ$ which is 
then progressively reduced in the ordered phase for finite $\tga$ due to the gap opening as shown in Fig.~\ref{fig:sus}. The value of $\chi(\bQ)$ also determines the effective intersite exchange constant J (Table \ref{tbl:matel}) for each $\tga$ which also fixes
the magnon energy scale $\omega_0$. We can see in Fig.~\ref{fig:sus} that the maximum in $\chi(\bq)$ remains at $\bq=\bQ$ for
all on-site exchange couplings $\tga$ considered and thus the commensurate AF order at this wave vector is consistent with
the chosen model parameters.\\

These static modifications of normal and anomalous conduction electron spectrum are to be complemented by the dynamic
magnon self energy effects which are the main topic of this work. The magnon excitations themselves have been kept as 
simple as possible for the AF structure with just the effective n.n. interaction included. Their dispersion according to Eq.~(\ref{eq:magdisp}) is shown in the inset of Fig.~\ref{fig:magnon-disp}(a) and it exhibits the periodicity $\omega_\bq=\omega_{\bq+\bQ}$ corresponding to the twofold degeneracy of magnon modes when downfolded to the AFBZ. The main figure shows the corresponding magnon DOS for increasing exchange anisotropy $\De$ and it is peaked at the upper X-point frequency. The $\bq$- dependence of the coherence factor $W_\bq$ (Eq.~(\ref{eq:cohfac})) which enters the magnon interaction with conduction electrons described by Eq.~(\ref{eq:Hcm}) is seen in Fig.~\ref{fig:magnon-disp}(b). It has a maximum at the M-point (equivalent to $\Gamma$ in the AFBZ), however in the self energy this is compensated by the magnon DOS which approaches the minimum at the M-point energy. Conversely the magnon DOS has its peak at the X-point energy  where the coherence factor is only unity. The dependence of the maximum magnon frequency on exchange coupling $\tga$ is shown in the inset of (b).\\

The momentum dependence of the self energy in the BZ as resulting  from the conduction electron coupling to magnons is shown in Fig.~\ref{fig:self-disp},  with their systematic evolution as function of coupling size $\tga$. The real part (black lines) which is responsible for the quasiparticle renormalisation shows a distinct sign change at the M-point  when $\tga$ increases from $0.2$ to  $0.8$. Synchronically the imaginary part which determines the quasiparticle lifetime peaks at the M point when the sign change appears and then becomes more evenly distributed for higher $\tga$ values. The main peak of the normal spectral function $\rho_0(\omega)$ comes from the states around the X-point. We also show a BZ plot of the change of quasiparticle dispersion in Fig.~\ref{fig:quasidisp} as modified by the real part of the self energy and the corresponding broadening due to imaginary part  line which shows clearly the renormalisation of the dispersion due to the coupling to magnons.  We plotted these curves in the original extended BZ of Fig.~\ref{fig:quasidisp} (full lines). The AFBZ (dashed lines) is rotated and has half the size. The $\Gamma$M midpoint X' is now an AFBZ boundary point and $\Gamma$, M become equivalent. Therefore the plots are symmetric with repsect to X'.
 \\

%
\begin{figure}
\includegraphics[width=0.9\columnwidth]{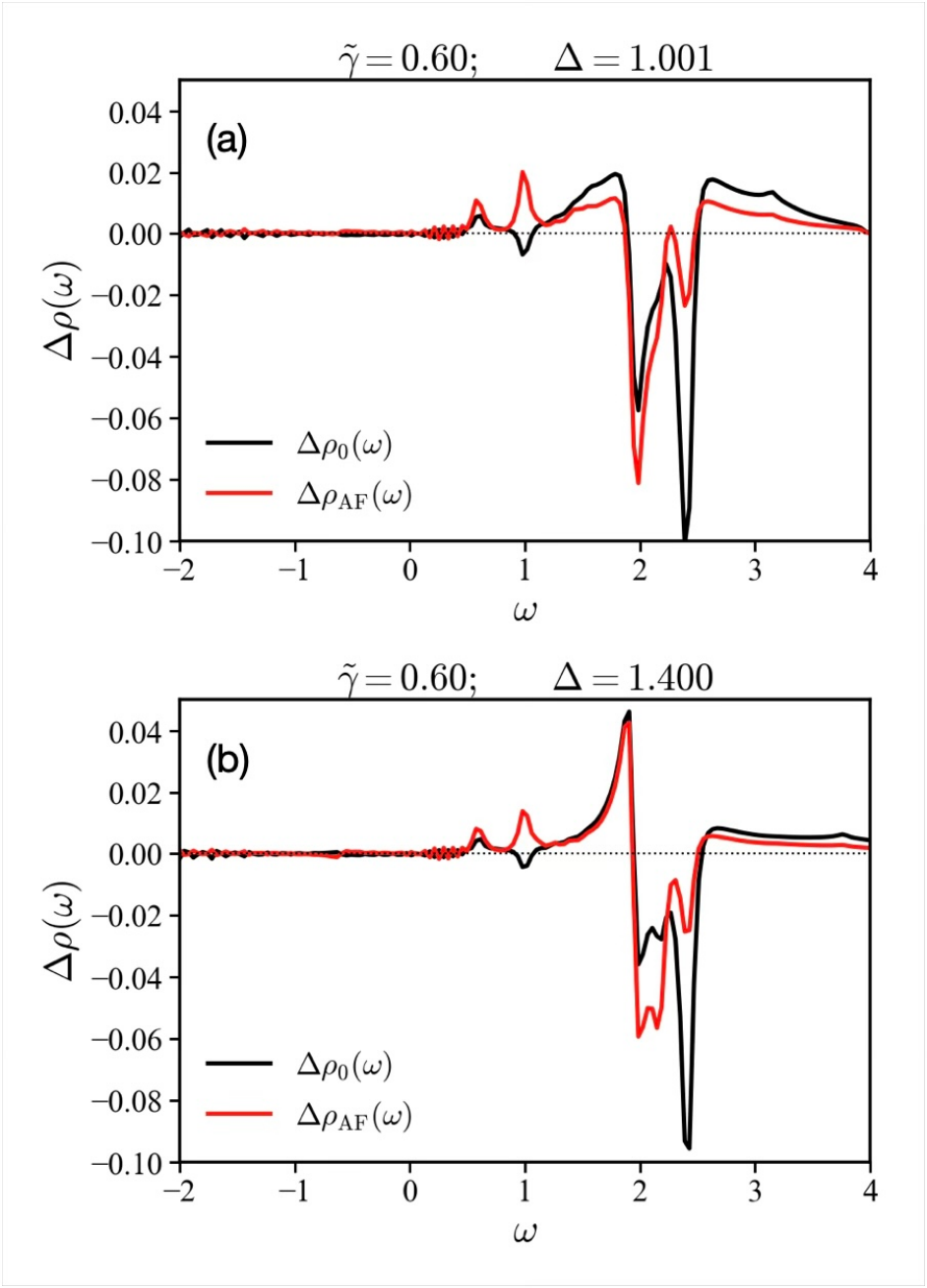}
\caption{Difference spectra $\Delta\rho_0$ (a) and $\Delta\rho_{AF}$ for $\tga=0.6$ for small (essentially absent) and large exchange anisotropy $\Delta$ (b) . The concomitant higher energy magnon spectrum (Fig.\ref{fig:magnon-disp}(a) leads to a suppression of the magnon sideband features in (b) as compared to (a) and also Fig.~\ref{fig:fullspec}(d) ($\Delta=1.1$). }
\label{fig:specdelta}
\end{figure}
%

Finally we discuss how the spectral functions visible in the STM tunnelling experiments (Eqs.~(\ref{eq:STM},\ref{eq:totDOS1},\ref{eq:STMspectra})) are modified by the band reconstruction caused by the static AF order parameter as well as the dynamic magnon coupling leading to the self energy. It should be said from the outset that the calculations have shown that both contributions to the STM spectrum are important. The former static correction was already discussed before in connection with Fig.~\ref{fig:disp-DOS} and it was found that a sizeable coupling $\tga$ is necessary to get visible modifications of the bare normal conduction band spectrum at $\bq=0$, whereas the anomalous spectrum which is zero in that case appears immediately for finite $\tga$ at the AF wave vectors $\bq=\pm\bQ$. The Fig.~\ref{fig:fullspec}(a) shows in one panel both the normal $\rho^{(0)}_0(\omega)$ and anomalous part $\rho^{(0)}_{AF}(\omega)$ without self energy correction (dashed lines, equivalent to  Fig.~\ref{fig:disp-DOS}) whereas the full lines include the self energy corrections given by the  corresponding $\rho_{0}(\omega)$ and $\rho_{AF}(\omega)$ spectra. On the absolute scale the corrections are concentrated around the  main peak of the spectral functions. A clearer picture is provided when we form the differences $\Delta\rho_0=\rho_0-\rho^{(0)}_0$ and $\Delta\rho_{AF}=\rho_{AF}-\rho^{(0)}_{AF}$ which are shown in  Fig.~\ref{fig:fullspec}(b). Independent of the coupling strength $\tga$ it is found that the coupling to the magnons reduces the height of the main spectral peaks (region of full blue arrow) and at the same time shifts the spectrum to  lower and higher energies by developing `sidebands' (indicated by dashed blue arrows). These sidebands are caused by the magnon absorption and emission part $(\pm\omega_\bq)$ in the self energy of Eq.~(\ref{eq:selfnum}). Their overall shape is determined by the magnon DOS and the momentum dependence of the coherence factors in Fig.~\ref{fig:magnon-disp}. This is also supported by Fig.~\ref{fig:specdelta} which, together with Fig.~\ref{fig:fullspec}(d) show the dependence of the difference spectra $\Delta\rho_{0},\Delta\rho_{AF}$ on the exchange anisotropy. The increase of the latter from $\Delta=1.001$ to $1.1$ and then $1.4$ shifts the magnon DOS to higher energies (Fig.~\ref{fig:magnon-disp}(a)). This suppresses the self energy
due to virtual magnon excitations and hence leads to a gradual suppression of sideband features in  Fig.~\ref{fig:fullspec}(d) and Fig.~\ref{fig:specdelta}(b) with increasing $\Delta$.  Therefore we conclude that aside from the band reconstruction by the static AF local moment order the dynamic magnon scattering leaves a visible imprint in the STM tunnelling spectrum of conduction electrons, both in normal and anomalous parts. Experimentally these effects should be more accessible in the latter anomalous part at $\pm\bQ$ since the background noise present at  \bq=0 (normal part) due to long wavelength inhomogeneities is not present at the finite AF wave vector.\\

\section{Summary and conclusion}
\label{sec:conclusion}

In this work we have focused on the influence of AF magnetic order of localised moments and their magnon excitations on the tunnelling current spectral functions in an STM setting. It was shown before for the compound \GR~both experimentally \cite{spethmann:24} and theoretically \cite{akbari:25} that, although the AF order is due to strictly localised 4f electron moments, nevertheless a clear signal of their ordering at the magnetic wave vector $\pm\bQ$ appears in the {\it conduction} electron tunnelling current. This was interpreted as a result of the
exchange coupling of the dual system, leading to a reconstruction of conduction bands and hence to an influence in the STM spectrum.

It remained unclear, however, whether the static description of the magnetic order is sufficient to interpret the STM spectrum,  in particular the anomalous part at the ordering vector. In fact its shape with frequency or bias voltage for such dual systems has
so far not yet been investigated in detail experimentally.
 In this work we have given a comprehensive theoretical discussion of the behavior to be expected from the inclusion of magnon scattering. We have restricted to a model system with simple AF local moment order
and a concomitant simple magnon spectrum and interaction Hamiltonian with conduction electrons. 

We have derived the self energy of conduction electrons due to scattering from magnons in perturbation theory with respect to the onsite exchange scattering. The self energy was incorporated into the derivation of the real-space Greens function and its Fourier transform on the same footing with
the band reconstruction and the two contributions to the spectrum were calculated. We found for a large range of exchange coupling strengths a clear evidence that the dynamic magnon self energy leads to sizeable correction in the spectrum contributions. They
consist of i) a reduction of the main peak height in both parts and ii) a development of sidebands due to the magnon scattering.
Their features change with the size of exchange coupling and magnon dispersion. We conclude that an interpretation of the  experimental STM spectra in such exchange coupled dual system with localised moment order must necessarily include the contributions due to conduction electron - magnon scattering.



\begin{acknowledgments}
A. A. acknowledges financial support from the Beijing Natural Science Foundation under Grant No. IS25015.
\end{acknowledgments}

\bibliography{References}

@article{rotter:12,
	author = {Rotter, Martin and Le, Manh Duc and Boothroyd, Andrew T and Blanco, Jesus Angel},
	doi = {10.1088/0953-8984/24/21/213201},
	journal = {Journal of Physics: Condensed Matter},
	month = {may},
	number = {21},
	pages = {213201},
	publisher = {IOP Publishing},
	title = {Dynamical matrix diagonalization for the calculation of dispersive excitations},
	url = {https://doi.org/10.1088/0953-8984/24/21/213201},
	volume = {24},
	year = {2012}}

@article{fulde:85,
	author = {Peter Fulde and Michael Loewenhaupt},
	doi = {10.1080/00018738500101821},
	journal = {Advances in Physics},
	number = {5},
	pages = {589--661},
	publisher = {Taylor \& Francis},
	title = {Magnetic excitations in crystal-field split 4f systems},
	volume = {34},
	year = {1985}}

@article{srivastava:05,
	author = {Srivastava, Pooja and Singh, Avinash},
	doi = {10.1103/PhysRevB.72.224409},
	issue = {22},
	journal = {Phys. Rev. B},
	month = {Dec},
	numpages = {10},
	pages = {224409},
	publisher = {American Physical Society},
	title = {Hole and electron dynamics in a triangular-lattice antiferromagnet: Interplay of frustration and spin fluctuations},
	url = {https://link.aps.org/doi/10.1103/PhysRevB.72.224409},
	volume = {72},
	year = {2005}}

@misc{ozdemir:26,
	archiveprefix = {arXiv},
	author = {Servet Ozdemir and Mikhail Kashchenko and Kostya S. Novoselov},
	eprint = {2605.00202},
	primaryclass = {cond-mat.mes-hall},
	title = {Observation of single antiferromagnetic magnon modes in the tunnelling transistors of spin-1/2 Kitaev system {$\alpha$-RuCl$_3$}},
	url = {https://arxiv.org/abs/2605.00202},
	year = {2026}}

@article{richmond:70,
	author = {P Richmond},
	doi = {10.1088/0022-3719/3/12/004},
	journal = {Journal of Physics C: Solid State Physics},
	month = {dec},
	number = {12},
	pages = {2402},
	title = {An electron in a ferromagnetic crystal (the magnetic polaron)},
	url = {https://doi.org/10.1088/0022-3719/3/12/004},
	volume = {3},
	year = {1970}}

@book{majlis:07,
	address = {Singapore},
	author = {Norberto Majlis},
	publisher = {World Scientific},
	title = {{The Quantum Theory of Magnetism}},
	year = {2007}}

@article{revenda:25,
	author = {Revenda, Jan and Wohlfeld, Krzysztof and Chaloupka, Ji\ifmmode \check{r}\else \v{r}\fi{}\'{\i}},
	doi = {10.1103/PhysRevB.111.195125},
	issue = {19},
	journal = {Phys. Rev. B},
	month = {May},
	numpages = {20},
	pages = {195125},
	publisher = {American Physical Society},
	title = {Magnetic polarons due to spin-length fluctuations in {${d}^{4}$} spin-orbit {Mott} systems},
	url = {https://link.aps.org/doi/10.1103/PhysRevB.111.195125},
	volume = {111},
	year = {2025}}

@article{maeland:21,
	author = {M\ae{}land, Kristian and R\o{}st, H\aa{}kon I. and Wells, Justin W. and Sudb\o{}, Asle},
	doi = {10.1103/PhysRevB.104.125125},
	issue = {12},
	journal = {Phys. Rev. B},
	month = {Sep},
	numpages = {12},
	pages = {125125},
	publisher = {American Physical Society},
	title = {Electron-magnon coupling and quasiparticle lifetimes on the surface of a topological insulator},
	url = {https://link.aps.org/doi/10.1103/PhysRevB.104.125125},
	volume = {104},
	year = {2021}}

@article{binghao:02,
	author = {Bing-Hao, Xie and Hong-Biao, Zhang and Jing-Ling, Chen},
	doi = {10.1088/0253-6102/38/3/292},
	journal = {Communications in Theoretical Physics},
	month = {sep},
	number = {3},
	pages = {292},
	publisher = {The Chinese Physical Society},
	title = {{BCS} Ground State and {XXZ} Antiferromagnetic Model as {SU(2), SU(1,1)} Coherent States: An Algebraic Diagonalization Method},
	url = {https://doi.org/10.1088/0253-6102/38/3/292},
	volume = {38},
	year = {2002}}

@article{matsuyama:23,
	author = {Matsuyama, N. and Nomura, T. and Imajo, S. and Nomoto, T. and Arita, R. and Sudo, K. and Kimata, M. and Khanh, N. D. and Takagi, R. and Tokura, Y. and Seki, S. and Kindo, K. and Kohama, Y.},
	doi = {10.1103/PhysRevB.107.104421},
	issue = {10},
	journal = {Phys. Rev. B},
	month = {Mar},
	numpages = {10},
	pages = {104421},
	publisher = {American Physical Society},
	title = {Quantum oscillations in the centrosymmetric skyrmion-hosting magnet {${\mathrm{GdRu}}_{2}{\mathrm{Si}}_{2}$}},
	url = {https://link.aps.org/doi/10.1103/PhysRevB.107.104421},
	volume = {107},
	year = {2023}}

@article{nomoto:20,
	author = {Nomoto, Takuya and Koretsune, Takashi and Arita, Ryotaro},
	doi = {10.1103/PhysRevLett.125.117204},
	issue = {11},
	journal = {Phys. Rev. Lett.},
	month = {Sep},
	numpages = {6},
	pages = {117204},
	publisher = {American Physical Society},
	title = {Formation Mechanism of the Helical {Q} Structure in {Gd}-Based Skyrmion Materials},
	url = {https://link.aps.org/doi/10.1103/PhysRevLett.125.117204},
	volume = {125},
	year = {2020}}

@article{eremeev:23,
	author = {Eremeev, S. V. and Glazkova, D. and Poelchen, G. and Kraiker, A. and Ali, K. and Tarasov, A. V. and Schulz, S. and Kliemt, K. and Chulkov, E. V. and Stolyarov, V. S. and Ernst, A. and Krellner, C. and Usachov, D. Yu. and Vyalikh, D. V.},
	doi = {10.1039/D3NA00435J},
	issue = {23},
	journal = {Nanoscale Adv.},
	pages = {6678-6687},
	publisher = {RSC},
	title = {Insight into the electronic structure of the centrosymmetric skyrmion magnet {GdRu$_2$Si$_2$}},
	url = {http://dx.doi.org/10.1039/D3NA00435J},
	volume = {5},
	year = {2023}}

@article{bouaziz:22,
	author = {Bouaziz, Juba and Mendive-Tapia, Eduardo and Bl\"ugel, Stefan and Staunton, Julie B.},
	doi = {10.1103/PhysRevLett.128.157206},
	issue = {15},
	journal = {Phys. Rev. Lett.},
	month = {Apr},
	numpages = {6},
	pages = {157206},
	publisher = {American Physical Society},
	title = {Fermi-Surface Origin of Skyrmion Lattices in Centrosymmetric Rare-Earth Intermetallics},
	url = {https://link.aps.org/doi/10.1103/PhysRevLett.128.157206},
	volume = {128},
	year = {2022}}

@article{wood:23,
	author = {Wood, G. D. A. and Khalyavin, D. D. and Mayoh, D. A. and Bouaziz, J. and Hall, A. E. and Holt, S. J. R. and Orlandi, F. and Manuel, P. and Bl\"ugel, S. and Staunton, J. B. and Petrenko, O. A. and Lees, M. R. and Balakrishnan, G.},
	doi = {10.1103/PhysRevB.107.L180402},
	issue = {18},
	journal = {Phys. Rev. B},
	month = {May},
	numpages = {5},
	pages = {L180402},
	publisher = {American Physical Society},
	title = {Double-{$Q$} ground state with topological charge stripes in the centrosymmetric skyrmion candidate {${\mathrm{GdRu}}_{2}{\mathrm{Si}}_{2}$}},
	url = {https://link.aps.org/doi/10.1103/PhysRevB.107.L180402},
	volume = {107},
	year = {2023}}

@article{yasui:20,
	author = {Yasui, Yuuki and Butler, Christopher J. and Khanh, Nguyen Duy and Hayami, Satoru and Nomoto, Takuya and Hanaguri, Tetsuo and Motome, Yukitoshi and Arita, Ryotaro and Arima, Taka-hisa and Tokura, Yoshinori and Seki, Shinichiro},
	date = {2020/11/23},
	doi = {10.1038/s41467-020-19751-4},
	id = {Yasui2020},
	isbn = {2041-1723},
	journal = {Nature Communications},
	number = {1},
	pages = {5925},
	title = {Imaging the coupling between itinerant electrons and localised moments in the centrosymmetric skyrmion magnet {GdRu$_2$Si$_2$}},
	url = {https://doi.org/10.1038/s41467-020-19751-4},
	volume = {11},
	year = {2020}}

@article{spethmann:24,
	author = {Spethmann, Jonas and Khanh, Nguyen Duy and Yoshimochi, Haruto and Takagi, Rina and Hayami, Satoru and Motome, Yukitoshi and Wiesendanger, Roland and Seki, Shinichiro and von Bergmann, Kirsten},
	doi = {10.1103/PhysRevMaterials.8.064404},
	journal = {Phys. Rev. Mater.},
	month = {Jun},
	number = {6},
	pages = {064404},
	publisher = {American Physical Society},
	title = {{SP-STM} study of the multi-{Q} phases in {${\mathrm{GdRu}}_{2}{\mathrm{Si}}_{2}$}},
	url = {https://link.aps.org/doi/10.1103/PhysRevMaterials.8.064404},
	volume = {8},
	year = {2024}}

@article{hoffman:11,
	author = {Hoffman, Jennifer E},
	doi = {10.1088/0034-4885/74/12/124513},
	journal = {Reports on Progress in Physics},
	month = {nov},
	number = {12},
	pages = {124513},
	title = {Spectroscopic scanning tunneling microscopy insights into {Fe}-based superconductors},
	url = {https://dx.doi.org/10.1088/0034-4885/74/12/124513},
	volume = {74},
	year = {2011}}

@article{akbari:25,
	author = {Akbari, Alireza and Thalmeier, Peter},
	doi = {10.1103/gp2r-kzf3},
	issue = {24},
	journal = {Phys. Rev. B},
	month = {Jun},
	numpages = {16},
	pages = {245129},
	publisher = {American Physical Society},
	title = {Image of helical local moment magnetic order in the {STM} spectrum},
	url = {https://link.aps.org/doi/10.1103/gp2r-kzf3},
	volume = {111},
	year = {2025}}
\end{document}